\documentclass[showpacs,twocolumn,floats,superscriptaddress]{revtex4}
\usepackage{graphicx}
\usepackage{amsmath}

\newcommand{\bq}{\begin{equation}}
\newcommand{\ee}{\end{equation}} \newcommand{\fr}[2]{\frac{#1}{#2}}
\newcommand{\eps}{\varepsilon}

\begin{document}

\title{
Spectrum of the Andreev Billiard and Giant Fluctuations of the
Ehrenfest Time
 }
\author{P. G. Silvestrov}
\affiliation{Theoretische Physik III, Ruhr-Universit{\"a}t Bochum,
44780 Bochum, Germany}
 \affiliation{Instituut-Lorentz, Universiteit Leiden, P.O. Box
9506, 2300 RA Leiden, The Netherlands}

\begin{abstract}

The density of states in the semiclassical Andreev billiard is
theoretically studied and shown to be determined by the
fluctuations of the classical Lyapunov exponent $\lambda$. The
rare trajectories with a small value of $\lambda$ give rise to an
anomalous increase of the Ehrenfest time $\tau_E\approx
|\ln\hbar|/\lambda$ and, consequently, to the appearance of
Andreev levels with small excitation energy. The gap in spectrum
is obtained and fluctuations of the value of the gap due to
different positions of superconducting lead are considered.
\end{abstract}
 \maketitle

{\it Introduction.} --- The density of states in a metallic island
coupled to a superconductor is modified due to the proximity
effect~\cite{Tinkham}. The changes are most pronounced in the
vicinity of Fermi energy, where there opens a gap in the spectrum
of excitations. A ballistic chaotic normal region coupled to a
superconductor via the small constriction (NS interface) is called
the Andreev billiard~\cite{Kos95}. The spectrum of such billiards
was calculated a decade ago~\cite{Mel96} assuming the random
matrix description of the quantum dynamics. This approximation is
valid if the number of channels $N$ supported by the NS interface
is small. In spite of large
efforts~\cite{Lod98,Sil03L,Vav03,Inanc02,Jacq03,Korman04,Be05,Marlies0305},
there is currently no reliable calculation of the spectrum in the
most interesting case of the semiclassical Andreev billiard, when
the number of open superconducting channels scales with the Planck
constant as~$N\sim 1/\hbar$.

The properties of Andreev billiards are governed by the Andreev
reflection, when an electron trajectory is retraced by the hole
that is produced upon absorption of a Cooper pair at the NS
interface. At the Fermi energy $E_{F}$, the classical dynamics of
the hole is the time reverse of the electron dynamics, so that the
motion is strictly periodic. This periodicity, however, comes in
conflict with a quantum-mechanical evolution, since each Andreev
reflection adds an extra phase $\pi/2$ to the electron-hole wave
function. This phase is compensated by the difference of classical
actions of electron $(E_F+\eps)$ and hole $(E_F-\eps)$ along the
trajectory, where $\eps$ is the excitation energy. The longer the
interval $t$ is between Andreev reflections, the smaller energy
suffices to produce the missing phase $\eps=\hbar\pi/2t$. On the
other hand, the probability of particle trajectory to not touch
the NS-interface for a long time become exponentially small $\sim
e^{-t/t_D}$, with $t_D$ being a dwell time. This leads to the
prediction of an exponential suppression of the density of
states~\cite{Shom99,Mel96}
 \bq\label{soft}
\rho(\eps)\approx N({\hbar\pi}/{\eps^2
t_D})\exp\left({-{\hbar\pi}/{2\eps t_D}}\right).
 \ee
However, the density of states at small excitation energies,
corresponding to large times, is not captured by this formula. A
new time scale responsible for the spectrum at low
energies~\cite{Lod98} is the Ehrenfest time $\tau_E\sim |\ln
\hbar|/\lambda$, where $\lambda$ is a Lyapunov exponent in the
normal billiard. For longer times the initial quantum wave packet
$\Delta x\Delta p \sim \hbar$ acquires a macroscopic size due to
exponential~$\sim e^{\lambda t}$ divergency of trajectories. This
invalidates the trajectory based derivation that led
to~(\ref{soft}). It is believed that, below certain energy
$\eps_{\rm gap}\sim \hbar/\tau_E$, the density of Andreev states
vanishes exactly, $\rho(\eps<\eps_{\rm gap})\equiv 0$, but the
magnitude of the gap and the mechanism of its formation remained a
subject of controversial discussion~\cite{Sil03L,Vav03,Be05}.

At finite times, the value of the Lyapunov exponent $\lambda$
depends on the specific trajectory~\cite{ott}, leading to
fluctuations of the Ehrenfest time. Both the Andreev
spectrum~\cite{Mel96,Lod98,Inanc02,Sil03L,Vav03,Jacq03,Korman04,Be05,Marlies0305},
and the quantum to classical crossover in ballistic
transport~\cite{Ale96,Sil03,Tworz04,Jack04,Brow05} attracted a
great deal of interest recently, but the role of fluctuations of
the Ehrenfest time was never investigated. In this Letter we show
how the low energy density of states is determined by the large
Ehrenfest time fluctuations and solve the long standing problem of
the Andreev gap.

The distribution of finite time Lyapunov exponents is
parameterized as (here $\tau_0$  is, e.g., an averaged time
between bounces at the walls of normal billiard)
 \bq\label{spect}
P(\lambda,t)=\tau_0\exp[tF_t(\lambda)].
 \ee
In the case of chaos the limit $F_{t\gg
\tau_0}(\lambda)=F(\lambda)$ exists, with $F(\lambda)$ specific
for the dynamical model~\cite{endnoteMix}. The function
$F(\lambda)$ has a maximum at $\lambda=\lambda_0$, which is the
conventional self-averaging Lyapunov exponent [$F(\lambda_0)=0$].
Since all small values of $\lambda>0$ are present in the
distribution~(\ref{spect}), one may always find (rare)
trajectories with any large value of the Ehrenfest time. To build
a semiclassical eigenfunction of 2-dimensional billiard, however,
one needs a family of trajectories, all having the same interval
$t$ between the Andreev reflections. The explicit construction of
discrete Andreev levels from the tube of trajectories (whose
transverse Poincar\'{e} section is quantized via the
Bohr-Sommerfeld rule) is presented in Ref.~\cite{Sil03L}. The gap
in the spectrum is determined by the largest time for which the
number of trajectories is enough to form at least one eigenstate.
Quantitative counting of the number of trajectories is done with
the use of the concept of transmission band~\cite{Wirtz99,Sil03}.
This lead us to the expressions for the gap and the density of
states [(\ref{egap}) and(\ref{rho}) below] depending on
$F(\lambda)$, $t_D$, and a number $N$ of open channels in the NS
interface.

\begin{figure}
\includegraphics[width=6.1cm]{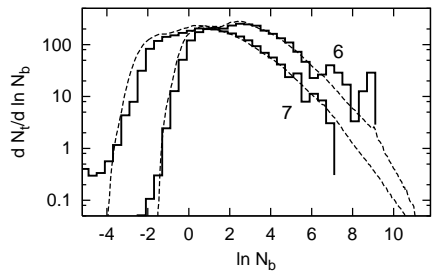}
 \vspace{-.2cm}
\caption{ Distribution of Andreev levels with
$\eps=\hbar\pi/12\tau_0$ and $\eps=\hbar\pi/14\tau_0$ over
transmission bands of different size, corresponding, respectively,
to the 6th and 7th iterations of the standard map averaged over
$200$ positions of the NS inteface. Dwell time $t_D=10$; kicking
strength $K=10$. The dashed lines show the theoretical
prediction~(\ref{distrib}), with the function $F$ found
numerically for 5 and 6 iterations. For $M=10^7$, the value
$\lambda=0$ would correspond to $N_b\approx 10^5$ in
Eq.~(\ref{Nband}). }
 \vspace{-.5cm}
\end{figure}

{\it Stroboscopic model.}--- All of the essential features of a
generic Andreev billiard are captured by the stroboscopic model of
the Andreev billiard, which
%we consider from now on. The model
was developed in Ref.~\cite{Jacq03} in order to use the advantages
of an open kicked rotator~\cite{Fyo00Oss03} for large scale
quantum-mechanical simulations. Here we present a semiclassical
solution of this model. First, the quantum kicked rotator, a
counterpart of the classical standard map~\cite{Chirikov79},
 \begin{equation}\label{kickrot}
p_{n+1}=p_n+{K I_0}\sin \theta_n/\tau_0 \ \ ,\ \
\theta_{n+1}=\theta_n+{\tau_0} p_{n+1}/I_0,
 \end{equation}
 is defined by the Floquet operator
 \bq\label{Fl1}
 U=\exp\left(i\fr{\hbar\tau_0}{2I_0}\fr{d^2}{d\theta^2}\right)
 \exp\left(i\fr{KI_0\cos \theta}{\hbar\tau_0}\right)
 .
 \ee
Here $I_0,\tau_0,$ and $K$ are the moment of inertia, the interval
between kicks, and the kicking strength, respectively. Next,
introduce the dimensionless Planck constant $\hbar_{\rm eff}=\hbar
\tau_0/I_0$. If $\hbar_{\rm eff}= 2\pi/M$ with integer $M$, the
coordinate and the momentum $\hat{p}=-i\hbar_{\rm eff} d/d\theta$
take discrete values $\theta_k=2\pi k/M$, $p_m=2\pi m/M$,
$k,m=1,2,...M$. The Floquet operator now becomes $M\times M$
matrix.

The electron and the hole components of the wave function of
Andreev kicked rotator span over the doubled $2M$-sites Hilbert
space with their evolution given by the normal ($U$) or conjugated
($U^\ast$) Floquet operators. The electron is converted into the
hole by reflection at the $N$-channel superconducting lead,
attached at $\theta_1<\theta<\theta_2$ ($\theta_2-\theta_1=\hbar
N$). This is done with the help of projection matrix $Q$ those $N$
only nonzero elements are $Q_{k,k}\equiv 1$, for $\theta_1<\hbar
k<\theta_2$. Andreev levels are found from
 \begin{equation}\label{Andremap}
{\cal U}\psi=e^{i\eps\tau_0/\hbar}\psi \ , \
{\cal U}=
\left(\begin{array}{cc}
(1-Q)U&-iQU^\ast \\
-iQU&(1-Q)U^\ast
\end{array}\right)
.
 \end{equation}
The classical limit corresponds to $M,N\rightarrow\infty$, while
the dwell time  $t_D=\tau_0 M/N$ is fixed. A classical particle at
any time $t=n\tau_0$ has a definite position either inside the
normal region or at the interface. A semiclassical quantization of
the map (\ref{Andremap}) requires a construction of the quantum
states having a similar property. A formal description of the wave
packet $\phi$, which is injected from the superconductor, stays
inside the billiard for $n-1$ kicks, and then hits the NS
interface is given by ($0<m<n$)
 \bq\label{eqUP}
Q\phi=\phi \ \ , \ \ QU^m\phi=0\ \ , \ \ QU^n\phi=U^n\phi .
 \ee
Provided such a solution is found, one easily builds $2n$
eigenfunctions with the eigenvalues~($r=0,...,n-1$)
 \bq\label{levels}
\eps_{nr}=\pm \hbar \pi \fr{2r+1}{2n\tau_0} \ , \
\psi=\left(\begin{array}{cc}
\sum_{k=0}^{n-1}U^k e^{-ik\eps\tau_0/\hbar}\phi\\
-\sum_{k=1}^{n}U^k e^{ik\eps\tau_0/\hbar}\phi
\end{array}\right).
 \ee
$\eps_{n0}$ with the largest possible $n$ constitutes a gap. Below
we always consider only the levels with $r=0$.
Equations~(\ref{eqUP}) and (\ref{levels}) are the analog of
adiabatic quantization, developed in ref.~\cite{Sil03L} for the
generic Andreev billiard.

Strictly speaking, Eqs.~(\ref{eqUP}) for any $n$ have no
solutions. However, for $\hbar\rightarrow 0$ where exists $\approx
N$ linearly independent wave packets satisfying~(\ref{eqUP}) with
practically any desired accuracy, $\phi^\dagger QU^m\phi\sim
e^{-1/\hbar}$. Finding the number of such solutions reduces to the
calculation of certain phase-space areas called transmission
bands~\cite{Sil03}.

%\begin{figure}
%\includegraphics[width=6.1cm]{PUMH67Sq.eps}
%%\includegraphics[width=6.5cm]{PUMH67Sq.eps}
%%\includegraphics[width=7.5cm]{PUMH67Sq.eps}
% \vspace{-.2cm}
%\caption{ Distribution of Andreev levels with
%$\eps=\hbar\pi/12\tau_0$ and $\eps=\hbar\pi/14\tau_0$ over
%transmission bands of different size, corresponding, respectively,
%to the 6th and 7th iterations of the standard map averaged over
%$200$ positions of the NS inteface. Dwell time $t_D=10$; kicking
%strength $K=10$. The dashed lines show the theoretical
%prediction~(\ref{distrib}), with the function $F$ found
%numerically for 5 and 6 iterations. For $M=10^7$, the value
%$\lambda=0$ would correspond to $N_b\approx 10^5$ in
%Eq.~(\ref{Nband}). }
% \vspace{-.5cm}
%\end{figure}

{\it Transmission bands.} --- We call the transmission band a
simply connected part of the phase-space area of the NS interface,
$\theta_1 < \theta <\theta_2, 0<p_{\rm eff}<2\pi$, each point
$\theta,p$ of which visits the interface at the $n$-th iteration
of the map (and do not visit it earlier). The image of the stripe
$\theta_1 < \theta <\theta_2$ after $n$ iterations is another long
and narrow (curved) stripe of a width $\sim e^{-\lambda t}N/M$
(see examples in Refs.~\cite{Tworz04,Jack04}). Phase-space
overlaps of the NS interface with its image, the transmission
bands, are the areas with approximately the shape of parallelogram
whose long and short sides have a length $\sim N/M$ and $\sim
e^{-\lambda t}N/M$. The number $N_b$ of (families of) Andreev
levels ~(\ref{levels}) supported by a single transmission band is
calculated as its area divided by $2\pi\hbar_{\rm eff}$,
 \bq\label{Nband}
N_b \approx ({N^2}/{M})e^{-\lambda t}.
 \ee
The total number of levels composed from trajectories having the
time $t=n\tau_0$ between hitting the NS interface, whose Lyapunov
exponent falls in the interval $d\lambda$, is~(\ref{spect})
 \bq\label{Ntime}
dN_t=NP(\lambda,t)e^{-{t}/{t_D}} (\tau_0/t_D)d\lambda.
 \ee
These levels originate from the many transmission bands of various
size: $N_t=\sum N_b$. We may use~(\ref{Nband}) to express
$\lambda$ through $N_b$ and to find the distribution of levels
over the sizes of the bands
 \bq\label{distrib}
\fr{dN_t}{d\ln N_b}=\fr{N^2\tau_0}{Mt}\exp\left\{ t
F_{t}\left(\fr{1}{t}\ln\fr{N^2}{MN_b}\right) -\fr{t}{t_D}\right\}.
 \ee
The distribution $dN_t/d\ln N_b$, found numerically for the
model~(\ref{Andremap}), is shown by the histogram in Fig.~1 for
times $t=6\tau_0$ and $t=7\tau_0$. The choice of $\hbar_{\rm eff}
=2\pi \times 10^{-7}$ introduces a quantum-mechanical scale in
classical area counting. Direct quantum-mechanical calculation of
energy levels in such an Andreev billiard would require
diagonalization of the matrix of the size $M=10^7$, which is
beyond the reach for existing computers.

The dashed lines in Fig~1 show the predicted distribution
[Eq.~(\ref{distrib})] with the numerically found function
$F_t(\lambda)$ for times $t=5\tau_0,6\tau_0$. The number of
iterations in the theoretical formula is reduced by $1$ because
the effective strong ($\sim K$) stretching of the phase-space
image of the NS interface starts only from the second iteration.

Even though the trajectories with $\lambda\approx 0$ are present
for both times shown in Fig~1, the largest transmission bands
expected from eq.~(\ref{Nband}) at the smallest values of the
Lyapunov exponent are absent~\cite{bending}. Also, the area of the
largest existing transmission band is smaller for a larger time.
This is the actual reason for the formation of the gap in the
excitation spectrum.

\begin{figure}
\includegraphics[width=6.4cm]{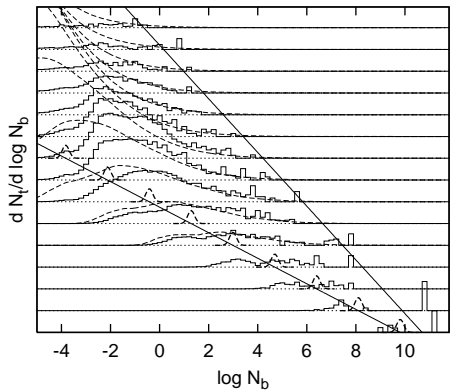}
 \vspace{-.2cm}
\caption{ The number of Andreev levels vs logarithm of the area of
transmission band $dN_t/d\ln N_b$ for the stroboscopic model with
$K=11$, $\hbar_{\rm eff} = 2\pi\times 10^{-7}$. The
superconducting lead  is attached at $\pi/10<\theta <3\pi/10$.
Histograms for different number of kicks $n=2-16$ are offset
vertically and multiplied by $2^n$. For $n\ge 6$, theoretical
distribution~(\ref{distrib}) is also shown (dashed line). The gap
amounts to $\eps_{\rm gap}=\hbar\pi/30\tau_0$ instead of
$\eps_{\rm gap}=\hbar\pi/14\tau_0$ expected from
Refs.~\cite{Sil03L,Vav03,Be05}. }
 \vspace{-.5cm}
\end{figure}

The gap in spectrum is given (\ref{levels}) by the {\it longest}
time for which there exist the transmission bands with $N_b>1$
 \bq\label{tmax}
\eps_{\rm gap}=\hbar\pi/2t_{\max} .
 \ee
To illustrate the mechanism of the creation of the gap we show in
Fig.~2 the distribution $dN_t/d\ln N_b$ for times $t/\tau_0=2-16$
(shifted vertically). Counting the small areas numerically is a
very time consuming-procedure~\cite{endnote}. Starting from the
10th iteration more than half of Andreev levels are missing due to
the errors in area counting. Still, the results for the largest
transmission bands are trustable for all times presented on the
figure. The existing estimates of the Andreev
gap~\cite{Sil03L,Vav03,Be05} neglected the fluctuations of the
Lyapunov exponents (shown by narrow peaks connected by a line).
The right solid line in Fig.~2 shows the (asymptotic) time
dependence of the largest transmission band area. The ratio of the
slopes of two lines give the suppression of the Andreev gap due to
the Ehrenfest time fluctuations.

{\it Density of Andreev states.} --- The area of transmission band
equation~(\ref{Nband}) increases exponentially for small values of
$\lambda$, while the number of trajectories with small $\lambda$
is exponentially small~(\ref{spect}). These two conflicting
effects allow one to find the value of the Lyapunov exponent
leading to the largest, for a given time, transmission band.
Equating via~(\ref{distrib}) the total number of expected levels
with $\lambda<\lambda_c$ to the area of a single
band~(\ref{Nband}) with $\lambda=\lambda_c$ we find
 \bq\label{eqLambda}
F(\lambda_c)=-\lambda_c+t_D^{-1}.
 \ee
The value of the gap now is
 \bq\label{egap}
\eps_{\rm gap}=\hbar\pi \fr{\lambda_c}{2\ln(N\tau_0/t_D)}.
 \ee
The value of $\lambda_c$ depends on details of the specific model.
It was shown in Ref.~\cite{SP04} that the derivative $F'(\lambda)$
has a maximum at $\lambda=0$ and that $F'(0)\le 1$. Since the only
maximum of the function $F$ is $F(\lambda_0)=0$ we obtain
 \bq\label{LambdaA}
\lambda_c<\lambda_0/2.
 \ee
Thus we found that a value of the Andreev gap is at least twice
smaller than predicted previously~\cite{Sil03L,Vav03,Be05}.

Semiclassical methods may be applied for a description of
eigenstates constructed from the trajectories with the Lyapunov
exponents
 \bq
\lambda < \lambda_{\rm max} =t^{-1} \ln({N\tau_0}/{t_D}).
 \ee
For $\lambda>\lambda_{\rm max}$, the number of levels per
transmission band became less than $1$. The number of Andreev
levels associated with the time $t$ may now be found by
integration of Eq.~(\ref{Ntime}) over the interval
$\lambda_c<\lambda<\lambda_{\rm max}$. If $\lambda_{\rm max}>
\lambda_0$ this allows one to recover the known result
eq.~(\ref{soft}), which is now valid for $\eps\ln (N\tau_0/t_D)>
\hbar\pi\lambda_0$. The smaller energies may exist only due to
$\lambda<\lambda_{max}<\lambda_0$, for which we found a novel form
of the density
 \bq\label{rho}
\rho(\eps)\approx \fr{N\tau_0}{\eps t_D F'(\kappa)}\exp\left[
 \fr{\hbar\pi}{2\eps}F(\kappa)- \fr{\hbar\pi}{2\eps t_D} \right] ,
 \ee
where $\kappa=({\eps}/{\hbar\pi}) \ln ({N\tau_0}/{t_D})$.
Equation~(\ref{rho}) is valid for
$\lambda_c<\eps\ln(N\tau_0/t_D)/\hbar\pi<\lambda_0$.

Formulas~(\ref{egap}) and (\ref{rho}) are the main results of this
Letter. They describe not only the stroboscopic model, but any
Andreev billiard with chaotic dynamics coupled to a superconductor
through the $N$-channel lead. (In this case, $\tau_0$ may be
replaced by the averaged time between bouncing of the billiard
walls. The precise value of $\tau_0$ is not important since
$N\sim\hbar^{-1}\gg t_D/\tau_0$.). For the model~(\ref{Andremap})
the semiclassical density consists of a series of
$\delta$-function peaks at $\eps=\eps_{nr}$~(\ref{levels}), and
Eq.~(\ref{rho}) describes the smoothed envelope of this
distribution. Such peaks in the Andreev spectrum were seen in
simulations of Refs.~\cite{Jacq03,Marlies0305}.

In Eq.~(\ref{egap}), we found the averaged value of the
superconducting gap. Fluctuations of $\eps_{\rm gap}$ are caused
by the variations in the position of NS interface. These
fluctuations result from the fluctuations of the area of largest
transmission band at a given time, which is shown in Fig.~3. Since
the area of (even the largest) transmission band decreases
exponentially with time while the number of the bands
exponentially increases, we expected that the correlation length
for fluctuation of the largest area should also decrease with
time. No such increased sensitivity is seen in Fig.~3.

{\it Discussion.} --- Among the total number of $M$ Andreev
levels, $M_q$ levels could not be described semiclassically,
 \bq
M_q\approx M^{1-\fr{1}{\lambda t_D}} \left(1 + \ln N/\lambda t_D
\right) t_D^{-\fr{2}{\lambda t_D}}\gg 1.
 \ee
These levels originate from the phase-space area of the NS
interface covered by small transmission bands, $N_b\le 1$.
References~\cite{Vav03,Be05} predict the vanishing of the density
of these levels for $\eps < \hbar\pi\lambda_0 /2\ln N$. Although
these papers do not provide a rigorous calculation of the Andreev
spectrum, we refer to these results as an indication that the
levels missing in the adiabatic quantization~\cite{Sil03L} do not
change the low energy density of states~(\ref{rho}) found in this
Letter.

\begin{figure}
\includegraphics[width=6.3cm]{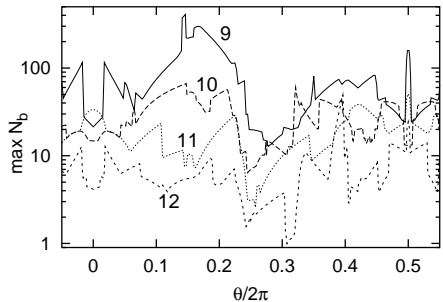}
 \vspace{-.2cm}
\caption{ The area of the largest transmission band (in units of
$\hbar =2\pi\times 10^{-7}$) for 9,10,11, and 12 iterations as a
function of the position of superconducting contact. Because of
the symmetries $\theta\rightarrow -\theta, \theta-\pi\rightarrow
\pi-\theta$, only a half of the $[0,2\pi]$ range of variation of
the argument is shown.}
 \vspace{-.5cm}
\end{figure}

The density of Andreev levels (\ref{rho}) and the Andreev gap
(\ref{egap}), which we found in this Letter, depend on the entire
distribution of finite time Lyapunov exponents $P(\lambda,t)$ [or
$F(\lambda)$~(\ref{spect})], not on the most probable Lyapunov
exponent $\lambda_0$, as was expected
previously~\cite{Lod98,Inanc02,Sil03L,Vav03,Jacq03,Korman04,Be05}.

The variations of the Lyapunov exponent leading to the
fluctuations of Ehrenfest time considered in this Letter are of
the order $\sim\lambda_0$. However, the corresponding variations
of the extent of the divergency of trajectories become
exponentially enhanced $\sim e^{\lambda t}$. So these are indeed
the giant fluctuations. The presented results are supported by the
classical numerical simulations. Verification of our findings in a
real experiment or in a quantum-mechanical
simulations~\cite{endnote1} remains a challenging problem.

Author acknowledges the assistance of A.~Tajic at the early stage
of the project. Discussions with M.V.~Fistul I.V.~Ponomarev and
C.W.J.~Beenakker are greatly appreciated. This work was supported
by the SFB TR 12 and by the Dutch Science Foundation NWO/FOM.


\vspace{-.35cm}

\begin{thebibliography}{99}

\bibitem{Tinkham} M. Tinkham, {\it Introduction to
Supercondactivity} (McGraw-Hill, New York, 1995).

\bibitem{Kos95} I. Kosztin, D.L. Maslov, and P.M.~Goldbart, Phys.\ Rev.\
Lett.\ {\bf 75}, 1735 (1995).

\bibitem{Mel96} J.A. Melsen, {\it et. al.},
%P.W. Brouwer, K.M.~Frahm, and
%C.W.J.~Beenakker,
Europhys.\ Lett.\ {\bf 35}, 7 (1996).

\bibitem{Lod98} A. Lodder and Yu.V.~Nazarov, Phys.\ Rev.\ B {\bf 58}, 5783
(1998).

\bibitem{Inanc02} \.{I}. Adagideli and C.W.J. Beenakker,
Phys.\ Rev.\ Lett.\ {\bf 89}, 237002 (2002).

\bibitem{Sil03L} P.G. Silvestrov, M.C. Goorden, and C.W.J. Beenakker,
Phys.\ Rev.\ Lett.\ {\bf 90}, 116801 (2003).

\bibitem{Vav03} M.G. Vavilov and A.I.~Larkin, Phys. Rev. B {\bf 67},
115335 (2003).

\bibitem{Jacq03} Ph. Jacquod, H. Schomerus, and C.W.J. Beenakker,
Phys.\ Rev.\ Lett.\ {\bf 90}, 207004 (2003).

\bibitem{Korman04} A. Korm\'{a}nyos, {\it et. al.},
%Z. Kaufmann, C.J.~Lambert,
%J.~Cserti,
Phys.\ Rev.\ B\ {\bf 70}, 052512 (2004).

\bibitem{Be05} C.W.J. Beenakker, Lect. Notes Phys. {\bf 667}, 131
(2005).

\bibitem{Marlies0305} M.C. Goorden, Ph. Jacquod, and C.W.J. Beenakker,
Phys.\ Rev.\ B\ {\bf 68}, 220501 (2003); {\it ibid}  Phys.\ Rev.\
B\ {\bf 72}, 064526 (2005).

\bibitem{Shom99} H. Schomerus and C.W.J. Beenakker,
Phys.\ Rev.\ Lett.\ {\bf 82}, 2951 (1999).

\bibitem{ott} E. Ott, {\em Chaos in Dynamical Systems} (Cambridge
University Press, Cambridge, 1993).

\bibitem{Ale96} I.L. Aleiner and A.I. Larkin, Phys.\ Rev.\ B\ {\bf 54}, 14423
(1996); O. Agam, I. Aleiner, and A. Larkin, Phys.\ Rev.\ Lett.\
{\bf 85}, 3153 (2000).

\bibitem{Sil03} P.G. Silvestrov, M.C. Goorden, and C.W.J. Beenakker,
Phys.\ Rev.\ B\ {\bf 67}, 241301 (2003).

\bibitem{Tworz04} J. Tworzyd\l o, A. Tajic, and C.W.J. Beenakker,
Phys.\ Rev.\ B\ {\bf 69}, 165318 (2004).

\bibitem{Jack04} Ph.~Jacquod and
E.V. Sukhorukov, Phys.\ Rev.\ Lett.\ {\bf 92}, 116801 (2004).

\bibitem{Brow05}  S. Rahav, P. W. Brouwer, Phys. Rev. Lett.,
{\bf 95}, 056806 (2005).

\bibitem{Wirtz99} L.~Wirtz, J.-Z.~Tang, and J.~Burgd\"{o}rfer,
Phys.\ Rev.\ B\ {\bf 59}, 2956 (1999).

\bibitem{endnoteMix}
In the case of mixed phase space, when there may exist stable
periodic orbits not touching the superconducting lead, the Andreev
gap disappears.

\bibitem{Fyo00Oss03} Y.V. Fyodorov and H.-J. Sommers, JETP Lett.\
{\bf 72}, 422 (2000);
A. Ossipov, T. Kottos, and T. Geisel, Europhys.\ Lett.\ {\bf 62},
719 (2003).

\bibitem{Chirikov79} B.V. Chirikov, Phys. Rep. {\bf 52}, 264 (1979).

\bibitem{bending} To describe the transmission bands formed by
the trajectories with $\lambda <\lambda_c$~(\ref{eqLambda}), one
has to consider dynamical folding of the phase space~\cite{SP04}.
The area of these bands is much smaller than predicted by
Eq.~(\ref{Nband}). Therefore, they do not appear in Fig.~1.

\bibitem{SP04} P. G. Silvestrov and I. V. Ponomarev,
nlin.CD/0409053.

\bibitem{endnote} Finding the distribution of Lyapunov exponents,
and consequently $F_n(\lambda)$, is much easier, as is seen in
Figs.~1 and 2.

\bibitem{endnote1}
Semiclassical quantization based on the transmission bands picture
was confirmed by Refs.~\cite{Jacq03,Marlies0305,Tworz04,Jack04}.
However, the value of the Planck constant $\hbar_{\rm eff}$
available in these simulations was to small to observe the
suppression of the gap~(\ref{egap}).

\end{thebibliography}
\end{document}